\pdfoutput=1
\documentclass[preprint,12pt]{elsarticle}

\usepackage{amssymb}
\usepackage{amsmath}
\usepackage{cite}
\usepackage{algorithmic}
\usepackage{graphicx}
\usepackage{textcomp}
\usepackage{multirow}
\usepackage{booktabs}
\usepackage{bbding}
\usepackage{xcolor}

\journal{arXiv}

\begin{document}

\begin{frontmatter}



\title{REHRSeg: Unleashing the Power of Self-Supervised Super-Resolution for Resource-Efficient 3D MRI Segmentation}


\author[1]{Zhiyun Song\fnref{fn1}}
\ead{zhiyunsung@gmail.com}
\author[2]{Yinjie Zhao\fnref{fn1}}
\ead{yinjie.zhao@monash.edu}
\author[3]{Xiaomin Li\fnref{fn1}}
\ead{lxm549496172@163.com}
\author[1]{Manman Fei}
\ead{feimanman@sjtu.edu.cn}
\author[1]{Xiangyu Zhao}
\ead{xiangyu.zhao@sjtu.edu.cn}
\author[1]{Mengjun Liu}
\ead{mjliu2020@sjtu.edu.cn}
\author[2]{Cunjian Chen}
\ead{cunjian.chen@monash.edu}
\author[2]{Chung-Hsing Yeh}
\ead{chunghsing.yeh@monash.edu}
\author[4]{Qian Wang}
\ead{qianwang@shanghaitech.edu.cn}
\author[1]{Guoyan Zheng}
\ead{guoyan.zheng@sjtu.edu.cn}
\author[3]{Songtao Ai\corref{cor1}}
\ead{aistss1024@sjtu.edu.cn}
\author[1]{Lichi Zhang\corref{cor1}}
\ead{lichizhang@sjtu.edu.cn}

\affiliation[1]{organization={School of Biomedical Engineering},%Department and Organization
            addressline={Shanghai Jiao Tong University}, 
            city={Shanghai},
            postcode={200030}, 
            country={China}}

\affiliation[2]{organization={Department of Data Science $\&$ AI, Faculty of Information Technology},
            addressline={Monash University},
            country={Australia}}

\affiliation[3]{organization={Department of Radiology},
            addressline={Shanghai Ninth People’s Hospital, Shanghai Jiao Tong University School of Medicine}, 
            city={Shanghai},
            postcode={200011}, 
            country={China}}

\affiliation[4]{organization={School of Biomedical Engineering},
            addressline={ShanghaiTech University}, 
            city={Shanghai},
            postcode={201210}, 
            country={China}}

\fntext[fn1]{Equally contribute to this paper.}
\cortext[cor1]{Corresponding authors.}

%% Abstract
\begin{abstract}
High-resolution (HR) 3D magnetic resonance imaging (MRI) can provide detailed anatomical structural information, enabling precise segmentation of regions of interest for various medical image analysis tasks. Due to the high demands of acquisition device, collection of HR images with their annotations is always impractical in clinical scenarios. Consequently, segmentation results based on low-resolution (LR) images with large slice thickness are often unsatisfactory for subsequent tasks. In this paper, we propose a novel \textit{Resource-Efficient High-Resolution Segmentation framework (REHRSeg)} to address the above-mentioned challenges in real-world applications, which can achieve HR segmentation while only employing the LR images as input. REHRSeg is designed to leverage self-supervised super-resolution (self-SR) to provide pseudo supervision, therefore the relatively easier-to-acquire LR annotated images generated by 2D scanning protocols can be directly used for model training. The main contribution to ensure the effectiveness in self-SR for enhancing segmentation is three-fold: 
(1) We mitigate the data scarcity problem in the medical field by using pseudo-data for training the segmentation model.
(2) We design an uncertainty-aware super-resolution (UASR) head in self-SR to raise the awareness of segmentation uncertainty as commonly appeared on the ROI boundaries.
(3) We align the spatial features for self-SR and segmentation through structural knowledge distillation to enable a better capture of region correlations.
Experimental results demonstrate that REHRSeg achieves high-quality HR segmentation without intensive supervision, while also significantly improving the baseline performance for LR segmentation.
\end{abstract}


% \begin{highlights}
% \end{highlights}

%% Keywords
\begin{keyword}


Magnetic resonance imaging \sep 
High-resolution segmentation \sep 
Self-supervised super-resolution \sep 
Uncertainty awareness \sep 
Knowledge distillation
\end{keyword}

\end{frontmatter}



\section{Introduction}
\label{sec1}
%% Labels are used to cross-reference an item using \ref command.

Magnetic resonance imaging (MRI) is widely used for diagnosis and monitoring due to its high precision in distinguishing between different types of soft tissue, while avoiding the risks associated with ionizing radiation exposure \citep{katti2011magnetic}. A critical task in computer-assisted diagnosis and intervention in MRI is the high-resolution (HR) 3D segmentation and reconstruction of regions of interest (ROI) within complex anatomical structures, which can facilitate subsequent procedures such as 3D printing and surgical planning \citep{QU2021101954}. 
In recent years, deep learning techniques have emerged as the leading approach in MRI segmentation, with 3D-based neural networks becoming the predominant method to achieve voxel-wise segmentation, as opposed to treating slices independently \citep{9446143}.  % \citep{milletari2016v, nnUNet, MedNeXt} 
Moreover, transformer-based techniques, which inherently capture global dependencies, have shown competitive performance, albeit at the cost of increased model complexity \citep{SHAMSHAD2023102802}. 
However, current methods for 3D MRI segmentation face challenges in clinical applications, where 2D scanning protocols are frequently employed to reduce acquisition time. Such protocols result in MR images with high in-plane resolution but lower inter-plane resolution, which cannot be directly processed by current segmentation models to achieve HR segmentation. 

Recently, several works have explored HR segmentation from low-resolution (LR) images \citep{patchfree1, patchfree2, GDN, video_patchfree}, presenting a potential solution to the aforementioned issues. These methods typically use downsampled images as the main input while producing HR segmentation directly, thereby avoiding the need for HR images during inference.
Some approaches also leverage image super-resolution as a proxy task to assist in capturing fine-grained features and ensuring that high-frequency details are preserved in the segmentation results \citep{patchfree1, patchfree2, DS2F}. However, these methods still rely on the availability of HR images and corresponding annotations for training, which are costly to obtain. On the one hand, prolonged scanning times for HR acquisition may cause patient discomfort and increase the risk of motion artifacts \citep{motionartifacts}; on the other hand, annotating HR images is considerably more labor-intensive and time-consuming. A potential solution to obtain HR segmentation with only LR images for training is to upsample the acquired LR images and interpolate the labels using morphology-based operators \citep{label_interpolate}. This strategy has been adopted in some datasets \citep{brats}, but it can lead to misaligned labels, particularly on the ROI boundaries.

\begin{figure}[t]
  \centering
  \includegraphics[width=.7\textwidth]{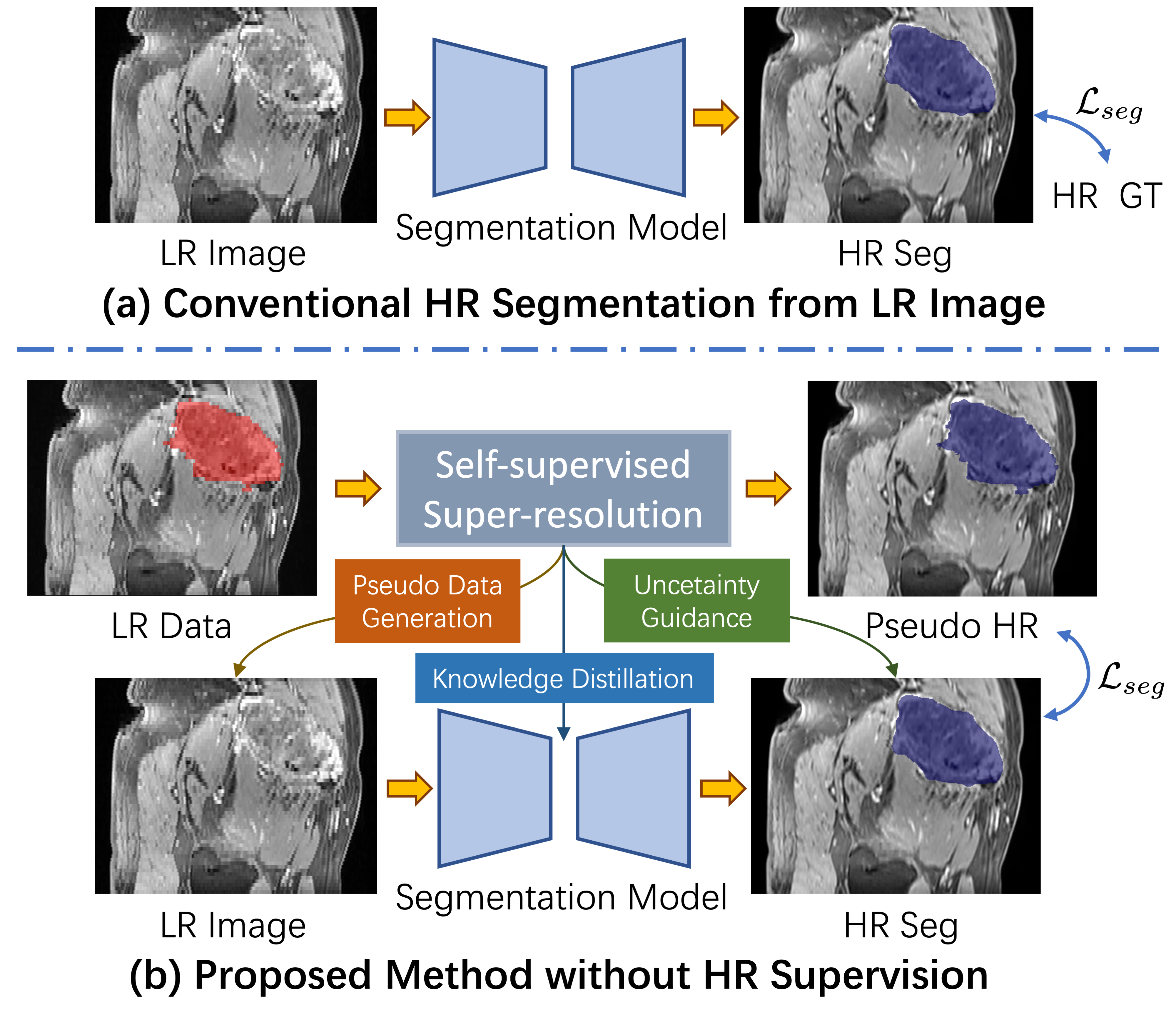}
  \caption{Comparison between \textbf{(a)} conventional high-resolution segmentation from low-resolution image method and \textbf{(b)} the proposed REHRSeg framework. Instead of using high-resolution annotations, we use a self-supervised super-resolution model to provide pseudo supervision for HR segmentation task, and further explore its capability to enhance the segmentation model from three different perspectives.} \label{graphic_abstract}
  \end{figure}

In this paper, we rethink current methods for HR segmentation from LR images and propose a novel \textit{Resource-Efficient High-Resolution Segmentation framework (REHRSeg)} for real-world clinical applications. 
As illustrated in Fig \ref{graphic_abstract}, conventional methods require HR annotated data, which are resource-intensive to collect. In contrast, REHRSeg replaces the supervision from real images with pseudo-HR data and addresses misalignment issues through a self-supervised super-resolution technique for annotated images. This strategy only requires LR data for training, making it more resource-efficient for practical use.

Furthermore, REHRSeg takes a step further in exploring super-resolution-assisted segmentation by investigating the capability of self-supervised super-resolution (self-SR) to enhance segmentation. This exploration begins by addressing data scarcity using pseudo-HR data generated by self-SR to expand the dataset for the segmentation task. Additionally, REHRSeg incorporates uncertainty-aware learning to improve ROI boundary recognition by utilizing the uncertainty map extracted from uncertainty-aware super-resolution (UASR) head, which highlights regions that are difficult to reconstruct. Moreover, REHRSeg aligns deep features in self-SR to capture correlated regions for segmentation. The key contributions of this work is as follows: 
\begin{itemize} 
\item We introduce a novel framework for high-resolution segmentation that can be trained without the need for high-resolution data.
\item We integrate a super-resolution prior to improve segmentation performance. Specifically, we: 
\begin{itemize} 
\item Employ super-resolution as a data generator to expand the training data for the segmentation task; 
\item Introduce an uncertainty-aware super-resolution (UASR) head within the super-resolution process to improve boundary recognition of segmentation task; 
\item Propose a distillation strategy to help the segmentation network capture deep correlations. 
\end{itemize} 
\item We conduct extensive experiments on both synthetic and real-world datasets, demonstrating that REHRSeg not only produces high-quality high-resolution segmentation but also significantly improves segmentation performance on low-resolution data.
\end{itemize}

\section{Related Work}
\label{sec:related}
\subsection{3D MRI Segmentation}
MRI segmentation aims to identify and delineate ROI in acquired MR images, playing a crucial role in computer-aided diagnosis and disease progression monitoring \citep{CLARKE1995343}. Recent advancements in deep learning technologies have led to significant breakthroughs in automatic MRI segmentation. Among the most widely adopted architectures are U-shaped CNN-based networks \citep{ronneberger2015u, DenseUNet, milletari2016v, MedNeXt}, with nnUNet \citep{nnUNet} emerging as a leading model. nnUNet automatically configures various datasets and focuses on dataset preprocessing and inference strategies, currently dominating the field of medical image segmentation. Furthermore, vision transformer (ViT) techniques have been applied to MRI segmentation, leveraging the transformer's ability to capture global and long-range dependencies \citep{nnformer, swinunetrv2, unetr++, transunet, ma2024segment}.
% Their flexibility in learning the entire feature map is particularly advantageous for tumor segmentation, accommodating varying shapes, sizes, and edge complexities.

Despite the high performance of these methods, they often require substantial computational resources for whole-volume segmentation of HR MRI images. Consequently, several approaches have aimed to develop efficient 3D MRI segmentation models with reduced resource demands by designing lightweight architectures.
% , including designing light-weight models, patch splitting, and downsampling. 
For example, ADHDC-Net \citep{lightweight1} uses decoupled convolution, dilated convolution, and attention-based refinement to minimize the number of parameters and operations in brain tumor segmentation.
Similarly, UNETR++ \citep{unetr++} employs efficient paired attention to reduce the complexity of self-attention computations and constrains the number of parameters by sharing weights between queries and keys. 
MISSU \citep{lightweight3} incorporates a self-distillation mechanism to refine local 3D features with multi-scale fusion blocks, which are removed during inference to ease computational burdens.
However, these lightweight models often struggle to capture rich features, limiting segmentation performance. Our approach addresses this limitation by utilizing the priors embedded in super-resolution models to enhance segmentation performance without increasing computational costs during inference.

\subsection{HR Segmentation from LR Image}
% To further reduce the demand for computational resources of processing the whole volume HR segmentation, one strategy orthogonal to efficient architecture as mentioned above is patch-splitting \citep{patchsplit1, patchsplit2}, which partition the whole image into smaller patches during the inference stage.
% However, the downfall is that it inevitably leads to chessboard issue \citep{DS2F} with highly discontinuous segmentation results and low inference speed \citep{ISDNet}.
Recently, several studies \citep{wang2021patch, patchfree2, ISDNet, video_patchfree, DS2F} have focused on achieving high-resolution (HR) segmentation from low-resolution (LR) images, primarily designed to efficiently capture global information and further reduce the computational cost, especially during inference.
For example, AR-Seg \citep{video_patchfree} reduces computational costs for video semantic segmentation by lowering the resolution for non-keyframes, while employing a cross-resolution fusion module to prevent accuracy degradation.
GDN \citep{GDN} replaces conventional down-sampling with a learnable procedure to reduce image resolution for real-time segmentation. 
UHRSNet \citep{shan2021uhrsnet} fuses local features extracted from image patches with global features from downsampled images. 
% This method effectively addresses the challenges posed by cropping and downsampling, leading to improved segmentation performance.

Super-resolution techniques have also been employed in this context to improve segmentation performance by capturing fine-grained details and providing richer semantic information. 
Lei \emph{et al.} \citep{S2Net} propose a framework for remote sensing images that simultaneously performs quadruple super-resolution and segmentation. 
PFSeg \citep{wang2021patch} integrates super-resolution as an auxiliary task for patch-free 3D medical image segmentation, and incorporates a fusion module for joint optimization.
ISDNet \citep{ISDNet} introduces super-resolution tasks as guides in the coarse-grained branches of the network, and fuses this information into fine-grained branches.
% This approach significantly accelerates inference while maintaining precise segmentation. 
%  Liu \emph{et al.} \citep{SRAL} introduce a general Super-Resolution-Assisted Learning (SRAL) framework to distill knowledge from super-resolution branches to segmentation branches.
DS2F \citep{DS2F} redesigns the segmentation and super-resolution framework by proposing a shared feature extraction module and a proxy loss for ROI recognition. 

While these methods achieve HR segmentation from LR images, they still rely on resource-intensive HR annotations during training. In contrast, our proposed REHRSeg framework only requires LR annotations for training, significantly reducing the annotation burden. Additionally, REHRSeg does not depend on HR images for super-resolution assistance, making it more practical for clinical applications where 2D scanning protocols are commonly employed.
 
\subsection{Self-supervised Super-resolution of MR Images}
Due to limitations in medical resources and scanning time, the acquisition of MR images employs a 2D scanning protocol in many clinical settings, resulting in low inter-plane resolution with large slice thickness \citep{synthsr, AFCM}. Therefore, super-resolution technology has been employed to enhance the resolution of such images, particularly in the inter-plane direction. Deep learning methods, especially CNN-based architectures, have become the dominant approach for MRI super-resolution. 
Conventional deep learning methods typically rely on fully supervised training \citep{DeepResolve, SAINT, SAINR}, requiring paired HR and simulated LR images.  However, collecting HR images in real-world clinical scenarios is significantly more challenging and costly than acquiring LR images, making it difficult to obtain sufficient paired training samples.

To address this, self-supervised training methods have been increasingly applied to super-resolution tasks.
These methods achieve super-resolution of MR images using only LR images for training. For example, Xuan \emph{et al.} \citep{xuankai} propose a VAE-based generative model trained on LR images, and synthesizes HR images from interpolated latent space to train the super-resolution model. SMORE \citep{SMORE} uses HR and LR data inherent in the high-resolution plane for training, restoring image quality by improving resolution and reducing aliasing. Wang \emph{et al.} \citep{SSR_ISBI} extend SMORE by introducing video frame interpolation as a preliminary task, and design an architecture based on implicit neural representation with a three-stage training protocol to refine the results. TSCTNet \citep{TSCNet} incorporates a cycle-consistency constraint to enable self-supervised learning for reducing the slice gap for MR images. 

In this work, we explore the capacity of self-supervised learning from annotated LR images in the field of MRI super-resolution to enhance and provide supervision for segmentation tasks, a direction not yet explored in prior super-resolution studies.

\begin{figure}[t]
\centering
\includegraphics[width=\textwidth]{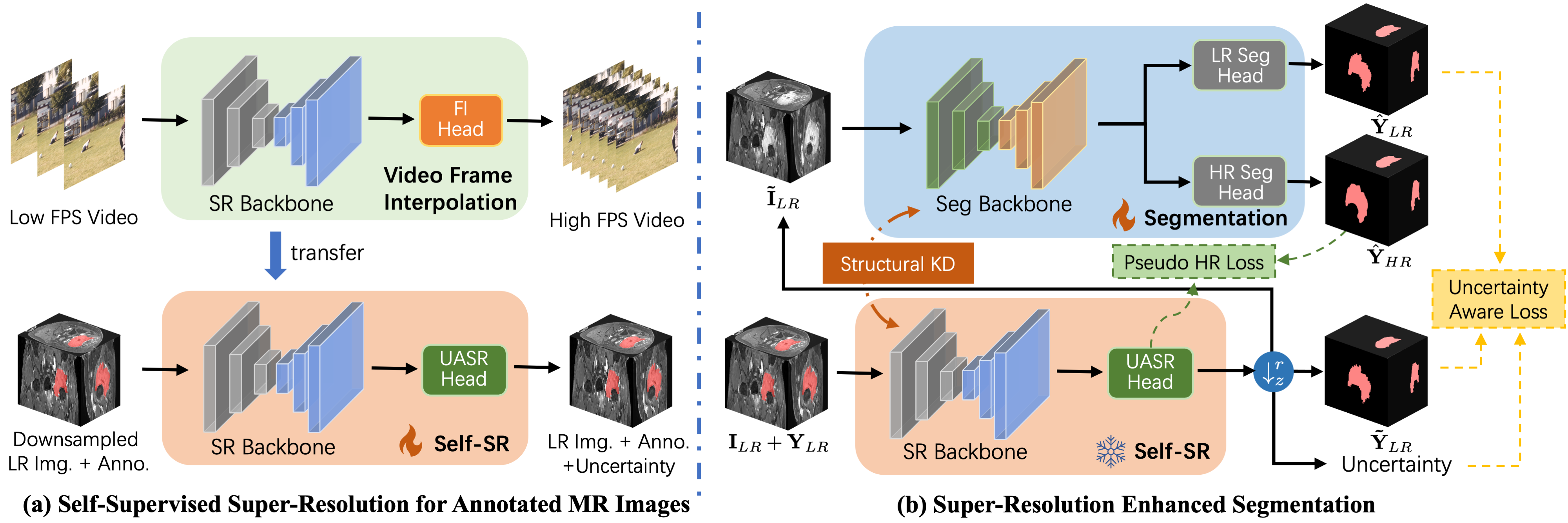}
\caption{Overall architecture of the proposed REHRSeg method, which includes two training stages: \textbf{(a)} Training of self-SR model for annotated MR images, and \textbf{(b)} Super-resolution guided segmentation. In the first stage, the self-SR model is initialized from the video frame interpolation (FI) model, and we introduce an uncertainty-aware super-resolution (UASR) head to ensure the awareness of uncertainty regions difficult to reconstruct. The self-supervision is achieved by learning from the mapping between the downsampled LR data and original LR data. In the second stage, self-SR can provide pseudo supervision for the segmentation task with the synthetic data. The uncertainty is used to help the segmentation model to recognize the blurred boundaries. We also introduce structural knowledge distillation (KD) between self-SR and segmentation models to help capture important correlations between regions.} \label{overall}
\end{figure}

\section{Method}
\label{sec:method}

\subsection{Overview}
The overall framework of REHRSeg is designed to produce high-resolution (HR) segmentation $\mathbf{\hat{Y}}_{HR}$, with an inter-plane resolution that is $r$ times higher than the original low-resolution (LR) image $\mathbf{I}_{LR}$ and its annotation $\mathbf{Y}_{LR}$. As illustrated in Fig. \ref{overall}, we begin by applying self-supervised super-resolution (self-SR) on the annotated MR images, generating coarse HR annotations $\mathbf{\tilde{{Y}}}_{HR}$ alongside the HR MR images $\mathbf{\tilde{{I}}}_{HR}$.
Next, the self-SR model and the pseudo HR data are leveraged to enhance the segmentation model, enabling simultaneous generation of both refined LR segmentation $\mathbf{\hat{Y}}_{LR}$ and HR segmentation $\mathbf{\hat{Y}}_{HR}$. This is achieved through three key components: 1) utilizing self-SR as a pseudo-data generator for segmentation, 2) incorporating an uncertainty-aware super-resolution (UASR) head in self-SR for additional boundary guidance, and 3) applying structural knowledge distillation from the self-SR model.

\subsection{Self-SR for Annotated MR images}
We introduce a self-supervised super-resolution (self-SR) method as a preliminary task in REHRSeg. This approach is built upon the principles of the most recent self-supervised MRI super-resolution method \citep{SSR_ISBI}, with modifications to support the super-resolution of annotated MR images. Given a LR image $\mathbf{I}_{LR}$ and its corresponding annotation $\mathbf{Y}_{LR}$, both with a spatial resolution of $a\times a \times c$, we assume that the frequency- and phase-encoding directions have identical spatial resolution. 
The super-resolution process is applied with a scaling factor of $r$, generating HR images with a spatial resolution of  $a\times a \times (c/r)$ in two main steps:
% , empowered by the most advanced video frame interpolation model \citep{flavr}.
First, both the image and annotation are interpolated to achieve isotropic voxel spacing, so that the spatial consistency can be ensured between training and inference stages. Our experiments show that the utilization of SMORE \citep{SMORE} as an alternative to traditional interpolation techniques, such as B-spline interpolation for images and nearest-neighbor interpolation for annotations, can further improve the final results.

In the second step, we generate the LR-HR pairs required for self-SR training by simulating slice separation along the $x$ axes. 
Specifically, the interpolated images are convoluted with a 1D Gaussian filter $h(x; r)$ as the slice profile with FWHM equal to $r$, followed by downsampling by a factor of $r$ along with the annotations. 
Unlike SMORE, we avoid introducing aliasing artifacts, as our self-SR model directly operates on $\mathbf{I}_{LR}$ instead of the interpolated one to reduce computational cost.
Additionally, the slice gap is not considered in our self-SR method, as many segmentation datasets lack this information. Notably, several SR methods such as SMORE\citep{SMORE}, DeepResolve\citep{DeepResolve}, and SAINR\citep{SAINR} also neglect the slice gap without experiencing significant performance degradation. 
In this scenario, the task can be considered as slice interpolation and deblurring for 3D data.

To accelerate training and ensure faster convergence, we leverage a pretrained model from the domain of video frame interpolation.
This strategy has been previously shown to be effective in self-supervised super-resolution \citep{SSR_ISBI}. 
Here we use the FLAVR \citep{flavr} model pretrained on Vimeo-90K \citep{Vimeo90k} to initialize the backbone of our self-SR model.
By applying the trained model to $\mathbf{I}_{LR}$ and $\mathbf{Y}_{LR}$, we obtain the estimated pseudo HR image $\mathbf{\tilde{I}}_{HR}$ and its annotation $\mathbf{\tilde{Y}}_{HR}$.
The pseudo data and the self-SR model are then used to enhance the segmentation model in the subsequent stage.

\subsection{Self-SR as Pseudo-data Generator}
We propose to utilize the self-SR model as a data generator to create pseudo-data for the MRI segmentation task, which is implemented in two distinct ways: data augmentation for LR segmentation and pseudo supervision for HR segmentation. 
Given the HR image $\mathbf{\tilde{I}}_{HR}$ and its annotation $\mathbf{\tilde{Y}}_{HR}$ produced by self-SR, we can simulate the LR data $\mathbf{\tilde{I}}_{LR}$ and $\mathbf{\tilde{Y}}_{LR}$ with a large slice thickness that matches the spatial resolution of the segmentation dataset, by blurring and downsampling the synthesized data as follows:

\begin{equation}
\begin{array}{l}
\mathbf{\tilde{I}}_{LR} = \{\mathbf{\tilde{I}}_{HR}*_{z}h(z; r)\}\downarrow^{r}_z, \\
\mathbf{\tilde{Y}}_{LR} = \{\mathbf{\tilde{Y}}_{HR}\}\downarrow^{r}_z,
\end{array}
\label{module}\end{equation}
where $*_z$ denotes 1D convolution along the $z$ axes, and $\downarrow^{r}_z$ represents the downsampling operator along the $z$ axes by a factor of $r$.
This process creates $r \times$ the amount of LR data compared to the original dataset, as the improved resolution in the synthesized data allows for additional data generation. This is accomplished by applying the downsampling operator while adjusting the corresponding starting index of the target image. Note that this can also be considered a novel form of data augmentation for LR segmentation, and the prior knowledge embedded within the self-SR process can be leveraged to further improve the segmentation performance with this additional data.

Furthermore, by generating the pseudo $\mathbf{\tilde{I}}_{LR}$-$\mathbf{\tilde{Y}}_{HR}$ pair via self-SR, the segmentation model is enabled to perform HR segmentation in parallel with LR segmentation task.
This is achieved by using an additional HR segmentation head consisting of a upsampling layer to the segmentation backbone, just before the original LR segmentation head.
During training, both tasks share the same input $\mathbf{\tilde{I}}_{LR}$, allowing for multi-task learning of LR and HR segmentation using pseudo-supervision from self-SR.

\subsection{Uncertainty-aware Learning with Self-SR Guidance}
We further utilize the self-SR model to provide uncertainty guidance for the segmentation task, with the goal of enhancing its performance with uncertainty-aware segmentation.
It is widely recognized that the tissues in medical images often overlap and blurriness occur on their boundaries, leading to high uncertainty in the delineation of these regions \citep{uncertainty_guidance_extend}.
This issues not only lead to unreliable segmentation region, but also cause poorly reconstructed regions for the super-resolution tasks.
Therefore, we propose to estimate uncertainty in the self-SR model by designing an uncertainty-aware super-resolution (UASR) head, which identifies the regions with high reconstruction error and provides guidance for the segmentation task.

The first step of uncertainty estimation with UASR involves the generation of intermediate results. We extract the 3D feature map $\mathcal{F}^{sr}_o \in \mathbb{R}^{C\times D\times H\times W}$ from the backbone of the self-SR model, where $D$, $H$, $W$ represent the slice depth, height, and width of the input LR images, respectively. The features along the depth dimension are concatenated into a 2D feature map, which is then processed to generate intermediate features $\mathcal{F}_m$ via Leaky ReLU activation and convolution.
These features are split along the channel dimension and merged back into the depth dimension, yielding $N$ intermediate images $\{P_I^j\}_{j=1}^N$ and annotations $\{P_Y^j\}_{j=1}^N$. The whole procedure can be expressed as:

\begin{equation}
\begin{array}{l}
\mathcal{F}_m = \text{LReLU}(\text{conv}(\text{merge}(\mathcal{F}^{sr}_o; 2; 1)), \\
P_{I}^j = \text{tanh}(\text{split}(\mathcal{F}_m; 1; 2)W_I^j+b_I^j), \\
P_{Y}^j = \text{split}(\mathcal{F}_m; 1; 2)W_Y^j+b_Y^j,
\end{array}
\label{multiple_pred}\end{equation}
where $\{W_Y^j, b_Y^j\}_{i=1}^N$ and $\{W_Y^j, b_Y^j\}_{i=1}^N$ are the 3D convolutional filters applied to the intermediate features.
The operators $\text{merge}(;a;b)$ and $\text{split}(;a;b)$ denote the merging of features along dimension $a$ into dimension $b$, and the splitting of features along dimension $a$ into dimension $b$, respectively.
Next, we obtain multiple attention maps by applying another set of convolutions, followed by a channel-wise softmax function:

\begin{equation}
P_{A}^j = \text{Softmax}(\text{split}(\mathcal{F}_m; 1; 2)W_A^j+b_A^j).
\label{multiple_attention}\end{equation}

\begin{figure}[t]
\centering
\includegraphics[width=.7\textwidth]{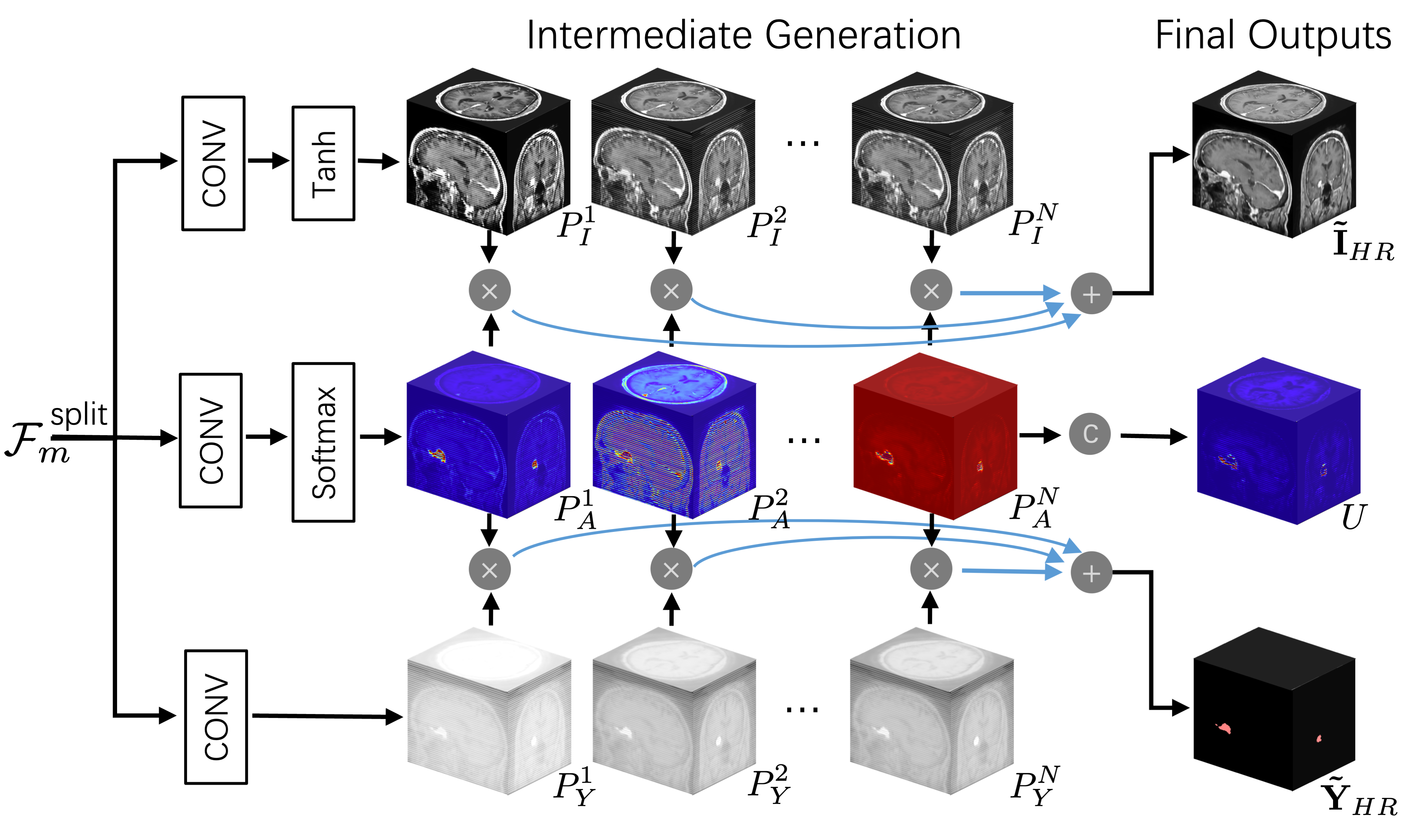}
\caption{Illustration of the proposed uncertainty-aware super-resolution (UASR) head for self-SR. The features from the last layer is processed through three independent branches, which produce the intermediate image maps, uncertainty maps, and segmentation maps. The final outputs for SR results are obtained by multiplying each intermediate generation with the uncertainty maps, followed by the addition operator.} \label{uncertainty_explain}
\end{figure}

Then, the attention maps are used to perform channel-wise selection for the intermediate images and annotations, and the final results are obtained via:
\begin{equation}
\begin{array}{l}
\mathbf{\tilde{I}}_{HR} = \sum_{j=1}^N P_A^j\otimes P_I^j, \\
\mathbf{\tilde{Y}}_{HR} = \sum_{j=1}^N P_A^j\otimes P_Y^j,
\end{array}
\label{final_output}\end{equation}
where $\otimes$ denotes element-wise multiplication. Similarly, the uncertainty map is also generated from the intermediate attention maps via:
\begin{equation}
U = \sigma(\text{conv}([P_A^1, P_A^2, \ldots, P_A^N])),
\label{uncertainty}\end{equation}
where $[\cdot]$ denotes concatenation along the channel dimension, and $\sigma$ is the sigmoid activation function. The modified loss function for the super-resolution task is used to ensure that the uncertainty map accurately reflects regions with high reconstruction error:
\begin{equation}
\mathcal{L}_{u}^{sr} = \frac{\mathcal{L}_{pix}^{sr}}{U} + \text{log}U,
\label{uncertainty_loss}\end{equation}
where $L_{pix}^{sr}$ is the original pixel-level loss function for the super-resolution task, which is defined as $L_1$ loss for the image. This modification ensures that the uncertainty map is automatically learned to represent the reconstruction error.
Note that we do not use the uncertainty map to regularize the super-resolution of annotations, which uses cross entropy and dice loss, to stabilize the learning process.

Since regions with high reconstruction error often correspond to boundaries of anatomical structures, we incorporate the uncertainty map to guide the segmentation task by introducing a segmentation loss regularization term:
\begin{equation}
\mathcal{L}_{u}^{seg} = \mathcal{L}_{pix}^{seg}\otimes (1 - \text{norm}(\{U\}\downarrow^{r}_z)),
\label{uncertainty_seg_loss}\end{equation}
where the uncertainty map extracted from UASR is first normalized to have its intensity range of $[0, 1]$ before employed as the weight map for the pixel-level segmentation loss.

\begin{figure}[t]
  \centering
\includegraphics[width=.7\textwidth]{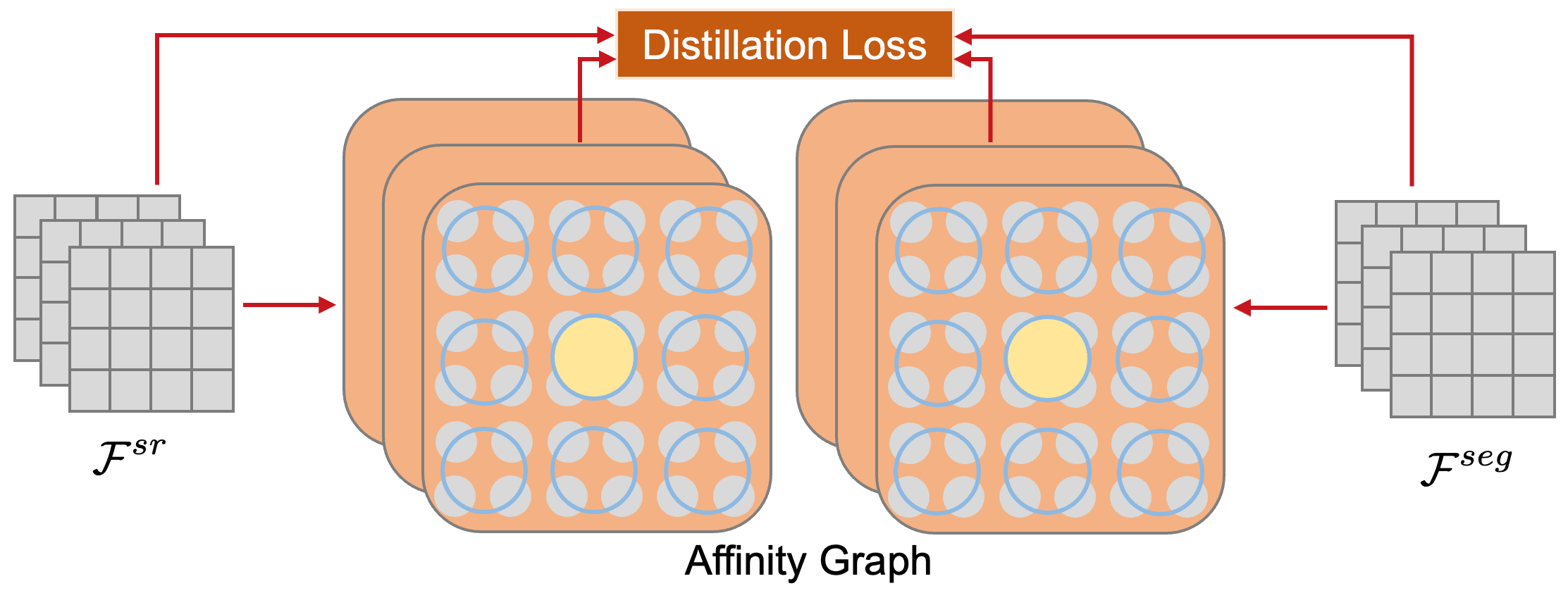}
\caption{Illustration of the structural knowledge distillation. The feature maps $\mathcal{F}^{sr}$ and $\mathcal{F}^{seg}$ are respectively extracted from self-SR and segmentation model. Knowledge distillation is performed by correlation distillation on the constructed affinity graph and spatial distillation on the feature maps.} \label{distill_explain}
\end{figure}

\subsection{Structural Knowledge Distillation}
We also propose to leverage knowledge distillation from the self-SR model to further enhance MRI segmentation. Given the disparity between image reconstruction and segmentation tasks, traditional distillation methods that constrain features with pixel-wise alignment only provide limited benefits. 
Instead, the structural correlation between regions shares similar patterns between the feature map produced by different tasks, which have also been confirmed in the previous work \citep{fei2024distillation}.
Therefore, the segmentation task has potential to benefit from the self-SR task by distilling the correlation patterns from it.
Here the feature maps $\mathcal{F}^{sr}$ and $\mathcal{F}^{seg}$ are extracted from the trained self-SR model and current segmentation model.
The shape of $\mathcal{F}^{sr}$ is aligned with $\mathcal{F}^{seg}$ with bilinear interpolation, resulting in feature maps of ${C'\times D'\times H'\times W'}$.
We model the spatial correlation between regions using a fully-connected affinity graph, where the nodes represent different spatial locations, and the edges denote their similarity.
In this graph, we aggregate the $\beta$ voxels in a local patch to represent the feature of each node, setting the granularity of the graph to $\beta$. Consequently, the affinity graph comprises $(D' \times H' \times W') / \beta$ fully connected nodes. We compute the similarity $a_{ij}$ between the $i$-th and $j$-th nodes by:
\begin{equation}
\begin{array}{l}
\mathcal{F}_p = \text{avgPool}(\mathcal{F}), \\
a_{ij} = \mathcal{F}_{p,i}^\top \mathcal{F}_{p,j}/(||\mathcal{F}_{p,i}||_2||\mathcal{F}_{p,j}||_2),
\end{array}
\label{similarity_graph}\end{equation}
where average pooling is applied to aggregate features before calculating similarity. This process is performed for both the self-SR and segmentation models, with $a_{ij}^{sr}$ and $a_{ij}^{seg}$ denoting the similarities from the self-SR and segmentation models, respectively. The loss for correlation distillation is defined as the squared difference between these similarity measures:
\begin{equation}
\mathcal{L}_{corr} = \frac{\beta}{D'\times W' \times H'}\sum_i \sum_j (a_{ij}^{sr}-a_{ij}^{seg})^2.
\label{distillation_loss_correlation}\end{equation}

Additionally, we align spatial features to highlight critical regions and improve segmentation performance. This is achieved through a cosine distance loss between the features from the self-SR model and the adapted segmentation features, as described in the previous work \citep{AM-RADIO}:
\begin{equation}
\mathcal{L}_{spatial} = \mathcal{L}_{cos}(h^{(v)}(\mathcal{F}^{seg}|\Theta^{(v)}), \mathcal{F}^{sr}),
\label{distillation_loss_spatial}\end{equation}
where $h^{(v)}(\cdot|\Theta^{(v)})$ is the student adaptor for feature vectors with parameters $\Theta^{(v)}$.

\subsection{Optimization}
The overall loss for training our segmentation model is composed of an uncertainty-aware segmentation loss, a pseudo HR segmentation loss, and a knowledge distillation loss:

\begin{equation}
\begin{aligned}
\mathcal{L}&=\mathcal{L}_{u}^{seg}(\mathbf{\hat{Y}}_{LR}, \mathbf{\tilde{Y}}_{LR}, U) + \mathcal{L}_{HR}^{seg}(\mathbf{\hat{Y}}_{HR}, \mathbf{\tilde{Y}}_{HR}) \\
&+\lambda (\mathcal{L}_{corr} (\mathcal{F}^{seg}, \mathcal{F}^{sr}) + \mathcal{L}_{spatial} (\mathcal{F}^{seg}, \mathcal{F}^{sr})),
\end{aligned}
\label{loss_g}\end{equation}
where $\lambda$ is used to balance the losses for segmentation and distillation.
$\mathcal{L}_{u}^{seg}$ is implemented with cross entropy loss weighted with uncertainty map, and $\mathcal{L}_{HR}^{seg}$ is implemented with cross entropy loss with dice loss.

\section{Experiments}
\label{sec:experiments}
\subsection{Datasets}
Our experiments are conducted on a public dataset with HR images and their annotations, as well as an in-house dataset with only LR labeled images.
Due to the difference in data availability, we calculate quantitative metrics on the public dataset for HR results, while providing only qualitative HR results for the in-house dataset. Both datasets undergo 5-fold cross-validation to ensure a robust and unbiased evaluation.

\subsubsection{Synthetic Public Dataset}
We first generate a synthetic LR MRI dataset from the publicly available Meningioma-SEG-CLASS dataset \citep{Meningioma-SEG-CLASS}, which contains pre-operative T1-CE (HR) and T2-FLAIR (LR) MR images from patients with pathologically confirmed Grade I or Grade II meningiomas. For our experiments, we extract 76 near-isotropic T1-CE images, each containing manually contoured tumors. Following standard practices in super-resolution techniques \citep{DeepResolve, AFCM, UniCOAL}, all images are first Gaussian-filtered before resampling to simulate the acquisition of thicker slices produced by 2D scanning protocols.
A downsampling factor of 4 is applied, and our method is completely blind to the original HR images and their annotations during training.

\subsubsection{In-house Dataset}
We also collect an internal dataset to evaluate the real-world application of our method.
The dataset contains 91 patients that underwent pelvic MRI exams using contrast-enhanced, fat-suppressed T1WI. 
The MR images are reconstructed with in-plane spatial resolution ranging from 0.42 to 0.74$mm^2$ and slice thickness between 3$mm$ and 6$mm$. The MR images were manually labeled to generate the gold standard for training and validation, which were performed by two radiologists using an annotation tool in the SenseCare research platform \citep{duan2020sensecare}. To standardize the data for segmentation model training, the in-plane resolution is resampled to 0.75$mm^2$, while the original thickness is kept unchanged during training, similar to \citep{qu2021surgical}.

\begin{table}[t]
\centering
\caption{\label{tab:lr_seg}
Quantitative results for low-resolution segmentation. 2D-based methods are also used to produce HR segmentation by slice-wise inference.}
\resizebox{\textwidth}{!}{%
\begin{tabular}{lccccccc}
\toprule
\multicolumn{1}{c}{\multirow{2}{*}{Method}}     & \multicolumn{4}{c}{Meningioma-SEG-CLASS}                                      &                      & \multicolumn{2}{c}{In-house}                 \\
\multicolumn{1}{c}{}                            & DSC (LR)         & HD95 (LR)         & DSC (HR)         & HD95 (HR)         &                      & DSC (LR)            & HD95 (LR)            \\ \midrule
\textbf{2D-based methods}                       &                   &                   &                   &                   &                      &                      &                      \\
SegFormer \citep{segformer}     & 0.7416$\pm$0.0846 & 29.01$\pm$22.13   & 0.6850$\pm$0.0879 & 42.52$\pm$9.80 &                      & 0.6150$\pm$0.0966    & 125.45$\pm$78.35    \\
% MaskFormer \citep{maskformer}   & 0.7873$\pm$0.0275 & 11.9$\pm$5.77     & 0.7723$\pm$0.0359 & 21.47$\pm$9.55 &                      & 0.7889$\pm$0.0607    & 57.77$\pm$77.30    \\
Mask2Former \citep{mask2former} & 0.8187$\pm$0.0546 & 21.05$\pm$18.91   & 0.8043$\pm$0.0606 & 23.49$\pm$13.84 &                      & 0.8444$\pm$0.0544    & 27.72$\pm$27.62    \\
nnUNet-2D \citep{nnUNet}        & 0.8137$\pm$0.0449 & 6.78$\pm$3.16 & 0.8077$\pm$0.0499 & 17.50$\pm$5.82 &                      & 0.8057$\pm$0.0528    & 28.55$\pm$12.41    \\
TransUNet-2D \citep{transunet}  & 0.7670$\pm$0.0743 & 18.94$\pm$16.28 & 0.7701$\pm$0.0696 & 21.76$\pm$6.84 &                      & 0.7589$\pm$0.0858    & 40.33$\pm$35.50    \\ \midrule
\textbf{3D-based methods}                       &                   &                   &                   &                   & \multicolumn{1}{l}{} & \multicolumn{1}{l}{} & \multicolumn{1}{l}{} \\
nnFormer \citep{nnformer}                                        & 0.8144$\pm$0.0274 & 14.31$\pm$12.91 & -                 & -                 &                      & 0.7957$\pm$0.0510    & 66.32$\pm$24.00    \\
TransUNet-3D \citep{transunet}  & 0.8067$\pm$0.0510 & 6.51$\pm$3.26 & -                 & -                 &                      & 0.7670$\pm$0.0723    & 33.01$\pm$23.41    \\
UNETR++ \citep{unetr++} & 0.7921$\pm$0.0650 & 17.61$\pm$15.57 & -                 & -                 &                      & 0.8761$\pm$0.0390    & 24.85$\pm$17.35    \\
nnUNet-3D \citep{nnUNet}        & 0.8155$\pm$0.0342 & 7.22$\pm$3.62     & -                 & -                 &                      & 0.8852$\pm$0.0456    & 15.00$\pm$10.10    \\
\textbf{REHRSeg}                                        & \textbf{0.8306}$\pm$0.0330 & \textbf{5.04}$\pm$2.93 & \textbf{0.8186}$\pm$0.0498 & \textbf{6.57}$\pm$2.38     &                      & \textbf{0.9041}$\pm$0.0308    & \textbf{9.41}$\pm$6.00    \\ \bottomrule
\end{tabular}
}
\end{table}

\begin{figure}[ht]
  \centering
  \includegraphics[width=\textwidth]{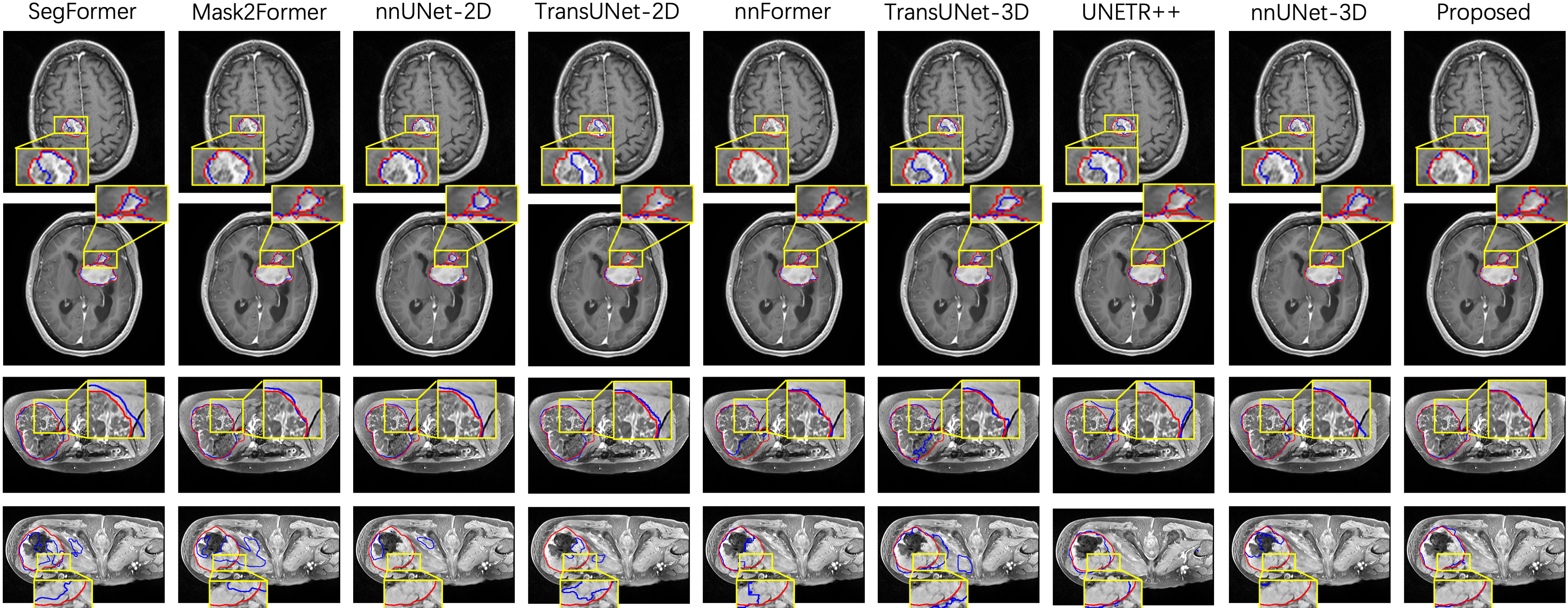}
  \caption{Qualitative results for low-resolution segmentation on Meningioma-SEG-CLASS dataset (the first two rows) and an in-house pelvic tumor dataset (the last two rows). The \textcolor{red}{red} lines denote the ground truth and the \textcolor{blue}{blue} lines denote the predictions.} \label{lr_seg_all}
  \end{figure}

\subsection{Implementation Details}
We utilize a downsampling factor (DSF) of 4 for both datasets, and initialize our self-SR backbone with the FLAVR model pretrained on $4 \times$ video frame interpolation. To stabilize the training of the UASR head, we only include the uncertainty-guided loss during the final 20,000 iterations, following an initial fine-tuning phase of 130,000 iterations with the batch size of 64. For structural distillation, we aggregate only in-plane features with a granularity $\beta$ set to $1 \times 2 \times 2$. The spatial distillation process employs a convolution layer following the previous work \citep{GAN_compression}, since experiments show that more complex distillers, such as those in \citep{AM-RADIO}, do not improve segmentation performance in this context.
The segmentation network is initialized using the default setting of nnUNet-3D \citep{nnUNet}, and its data augmentation strategy is applied throughout our experiments. To allow the segmenter to produce HR results, we interpolate the features before the last layer and use two convolution with a ReLU activation in between, which is light-weighted to produce the final results. All the experiments are performed on an NVIDIA RTX A6000 GPU. 

\subsection{Enhanced LR Segmentation}
In this section, we evaluate whether the self-SR model can provide effective guidance for the segmentation model, by inspecting whether its performance is improved for LR segmentation.
We also conduct experiments for HR segmentation on the Meningioma-SEG-CLAS dataset, as 2D-based segmentation methods can be directly applied to cross-sectional slices extracted from HR volumes. It is important to note that these 2D-based methods use HR images as inputs, whereas our method operates on LR images in the inference stage.
Quantitative results presented in Table \ref{tab:lr_seg} indicate that 2D-based methods generally underperform 3D-based methods for MRI segmentation, particularly in terms of the HD95 metric. Notably, nnUNet remains a strong baseline and outperforms several transformer-based methods, especially for the in-house dataset where the complex data distribution poses challenges for these methods to achieve satisfactory results.
This can be attributed to the limited amount of training data, which prevents transformer-based methods from fully leveraging their advantages.

Despite these challenges, REHRSeg significantly improves performance over baseline ($p<$ 0.05, HD95) and outperforms all the alternatives for both LR and HR segmentation. Visualization results in Fig. \ref{lr_seg_all} also demonstrate that REHRSeg excels in both tumor types, and notably enhances the nnUNet baseline in boundary recognition. Moreover, REHRSeg shows strong performance in difficult cases where other methods fail to accurately capture the tumor shapes, as illustrated in the fourth row of Fig. \ref{lr_seg_all}.

\begin{table}[t]
\centering
\caption{\label{tab:SR_results_quantitative}
Quantitative results for self-supervised super-resolution on the Meningioma-SEG-CLAS dataset.}
\begin{tabular}{lccc}
\toprule
Methods & PSNR & SSIM & DSC \\
 \midrule
B-spline & 25.12$\pm$0.41 & 0.9153$\pm$0.0120 & 0.9005$\pm$0.0689 \\
SMORE \citep{SMORE} & 29.73$\pm$0.63 & \textbf{0.9616}$\pm$0.0070 & 0.9302$\pm$0.0379 \\
FLAVR \citep{flavr} from scratch & 29.15$\pm$0.53 & 0.9532$\pm$0.0090 & 0.9340$\pm$0.0361 \\
\textbf{REHRSeg} & \textbf{29.86}$\pm$\textbf{0.56} & 0.9583$\pm$0.0083 & \textbf{0.9357}$\pm$0.0356 \\ \bottomrule
\end{tabular}
\end{table}

\begin{figure}[t]
  \centering
\includegraphics[width=\textwidth]{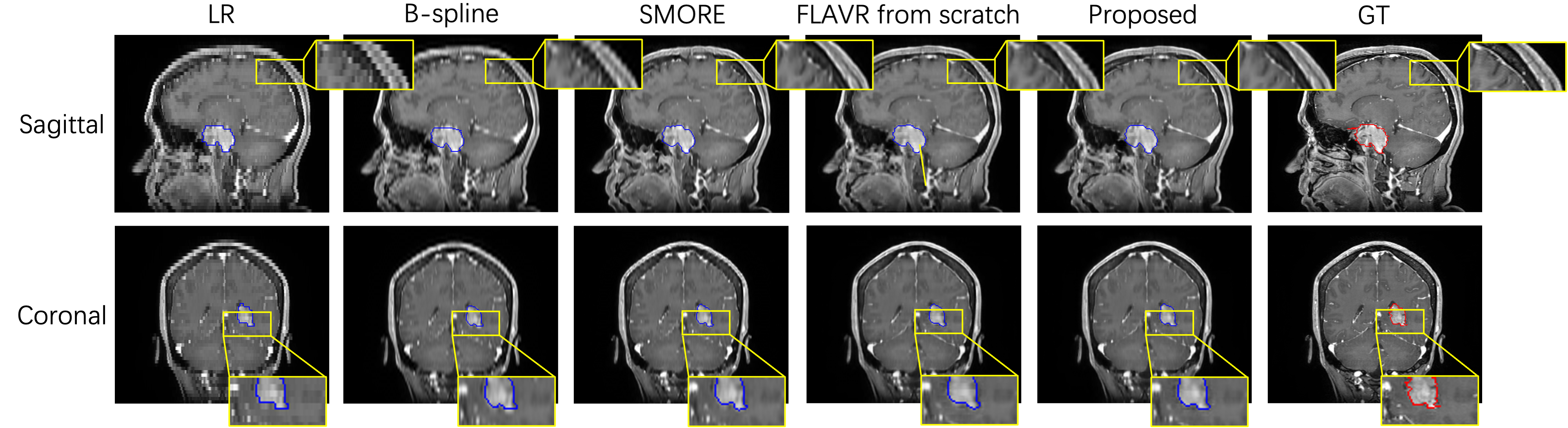}
\caption{Qualitative results for super-resolution of annotated MR images without ground truth.
} \label{sr_results}
\end{figure}

\subsection{Reconstructing HR Results Using LR Data}
Although REHRSeg is trained on LR images and their corresponding annotations, it allows the acquisition of coarse HR results via the proposed self-SR model. Additionally, the segmentation model is enhanced using these self-SR results to produce HR segmentation outputs. The evaluation of these results are further illustrated in the following sections.
\subsubsection{Super-resolution Results for Annotated Images}
We first evaluate the quality of our method for self-supervised super-resolution.
As there are no existing methods for the super-resolution of both images and labels, we implement three baseline methods, including: 1) B-spline interpolation for image and morphology-based interpolation for label \citep{label_interpolate} commonly used for producing spatially standardized data \citep{brats}, 2) a modified version of SMORE (2D) \citep{SMORE} with an auxiliary segmentation head, and 3) our model trained from scratch (\emph{i.e.}, without pretraining on video frame interpolation).

Quantitative results demonstrate that our method achieves the highest Peak Signal-to-Noise Ratio (PSNR) and Dice Score (DSC), although the Structural Similarity Index (SSIM) metric is slightly lower compared to SMORE. It is noteworthy that pretraining on a video frame-interpolation task significantly improves the performance of the self-SR model (from 29.15 to 29.86 in PSNR), which is also consistent with findings from previous work \citep{SSR_ISBI}.
The qualitative results shown in Fig. \ref{sr_results} illustrate the super-resolution outcomes for labels overlaid on the super-resolved images. Simple B-spline interpolation results in blurred images and misaligned labels. While SMORE offers notable improvements over B-spline interpolation, it still exhibits issues such as discontinuities on the surface of the pallium (sagittal view) and blurriness on the ROI boundaries (axial and coronal views). In contrast, our proposed self-SR method can preserve image-label correspondence with clear boundaries. Although some serration remains along the borders, further refinement is possible using a video-pretrained model. These precise super-resolution results serve as the foundation that the knowledge gained from our model can guide and enhance the following segmentation process.

\begin{table}[t]
\centering
\caption{\label{tab:SR_seg_results}
Quantitative results for HR segmentation from LR image on Meningioma-SEG-CLAS dataset, \dag means the method require auxiliary HR inputs, \ddag means the HR inputs only pass through a shallow network, and $\star$ means the model is trained with interpolated data.}
\begin{tabular}{lcccc}
\toprule
Methods & HR input & HR label & DSC & HD95 \\
 \midrule
PFSeg \citep{wang2021patch} & \XSolidBrush ${\dag}$ & \Checkmark & 0.7279$\pm$0.0664 & 15.25$\pm$9.23 \\
DS2F \citep{DS2F}  & \XSolidBrush  & \Checkmark & 0.7518$\pm$0.0538 & 12.88$\pm$9.09 \\
ISDNet \citep{ISDNet} & \Checkmark ${\ddag}$ & \Checkmark & 0.7090$\pm$0.0771 & 18.53$\pm$8.72 \\
nnUNet$\star$ \citep{nnUNet} & \XSolidBrush  & \XSolidBrush & 0.8035$\pm$0.0453 & 9.38$\pm$3.64 \\
\textbf{REHRSeg} & \XSolidBrush & \XSolidBrush & \textbf{0.8186}$\pm$0.0498 & \textbf{6.57}$\pm$2.38 \\ \hline
Fully Supervised & \Checkmark & \Checkmark & 0.8481$\pm$0.0414 & 8.92$\pm$2.65 \\ \bottomrule
\end{tabular}
\end{table}

\begin{figure}[t]
  \centering
  \includegraphics[width=\textwidth]{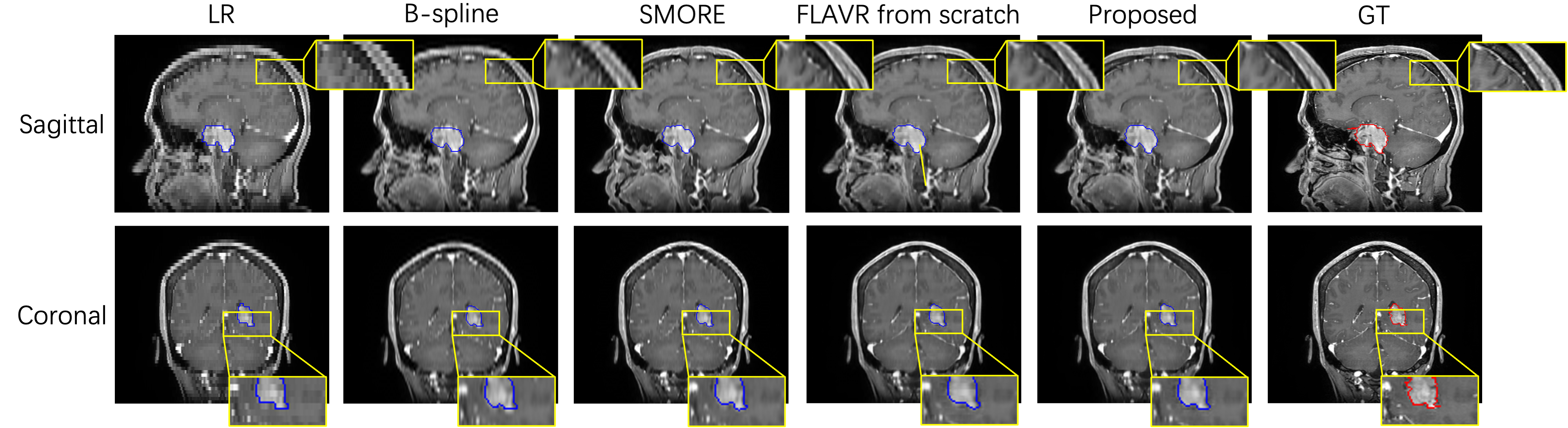}
  \caption{Qualitative results for HR segmentation given LR MR images. The \textcolor{red}{red} lines denote the ground truth and the \textcolor{blue}{blue} lines denote the predictions.
  } \label{hr_seg_results}
  \end{figure}

\subsubsection{HR Segmentation from LR Image}
We evaluate the performance of REHRSeg for HR segmentation from LR image in this section.
In addition to the previous works with 3D-based \citep{wang2021patch, DS2F} or 2D-based \citep{ISDNet} methods for HR segmentation, we also compare with the training protocol that uses the B-spline interpolated data defined in the previous section, which enables the trained model to produce HR segmentation results using upsampled LR images as inputs.
This will consume considerably more computational resources during training and inference due to the enlarged input size.
Quantitative results in Table \ref{tab:SR_seg_results} indicate that REHRSeg achieves top-ranked performance without any HR ground truth.
It is also noteworthy to find that our HD95 metric is even better than fully supervised method, which can be explained by our effective super-resolution assisted segmentation design.
The qualitative results for three different subjects in Fig. \ref{hr_seg_results} also support the reported metrics. For instance, 2D-based methods such as ISDNet exhibit better results in the sagittal view compared to other views, but their overall performance is inferior to 3D-based methods like PFSeg. Furthermore, training with interpolated data using a powerful segmentation framework provides a strong baseline for this task. However, the boundaries in these interpolated results are less precise, likely due to misalignment between the interpolated images and labels.
In contrast, REHRSeg provides the segmentation results whose boundaries are closest to the ground truth, even in challenging cases such as those from the coronal view, where other methods struggle to achieve satisfactory results. 

\begin{table}[t]
  \centering
  \caption{\label{tab:ablation_structure}
  Quantitative results for our ablation study on different architecture design.}
  \resizebox{\textwidth}{!}{%
  \begin{tabular}{lcccc}
  \toprule
                          & DSC (LR)      & HD95 (LR) & DSC (HR)      & HD95 (HR) \\ \midrule
  Baseline                & 0.8155$\pm$0.0342 & 7.22$\pm$3.62 & -             & -         \\
  +Pseudo SR Data         & 0.8213$\pm$0.0302 & 7.04$\pm$3.77 & 0.8007$\pm$0.0403 & 7.07$\pm$2.19 \\
  +Uncertainty Guidance   & 0.8231$\pm$0.0281 & 6.96$\pm$0.87 & 0.8027$\pm$0.0468 & 7.11$\pm$1.52 \\
  \textbf{+Knowledge Distillation} & \textbf{0.8306}$\pm$0.0330 & \textbf{5.04}$\pm$2.93 & \textbf{0.8186}$\pm$0.0498 & \textbf{6.57}$\pm$2.38 \\ \bottomrule
  \end{tabular}
  }
\end{table}

\subsection{Ablation Study}
\subsubsection{Effectiveness of Our Architecture}
We conduct an ablation study of each component of REHRSeg to evaluate their effectiveness, and report the quantitative results in Table \ref{tab:ablation_structure}.
The experiments start by the incorporation of pseudo data generated by our self-SR model to assist the segmentation model with extra data and an auxiliary task.
We can observe an improvement of the LR segmentation performance, and more importantly the availability of HR segmentation.
Next, the introduction of uncertainty guidance refines the segmentation model to be aware of the boundaries of ROI, indicated by more advanced segmentation results.
Our knowledge distillation strategy achieves the best performance, which significantly surpasses the baseline. This trend is also reflected in HR segmentation, although the addition of uncertainty guidance results in a slight decrease on the HD95 metric.

\begin{table}[t]
  \centering
  \caption{\label{tab:ablation_selfsr}
  Quantitative results for our ablation study on different self-SR methods.}
  \resizebox{\textwidth}{!}{%
  \begin{tabular}{lcccc}
  \toprule
                          & DSC (LR)      & HD95 (LR) & DSC (HR)      & HD95 (HR) \\ \midrule
  B-spline           & 0.8105$\pm$0.0421 & 7.52$\pm$3.69 & 0.7877$\pm$0.0590 & 7.21$\pm$2.21 \\
  SMORE              & 0.8226$\pm$0.0301 & 7.75$\pm$2.55 & 0.8068$\pm$0.0410 & 7.11$\pm$1.87 \\
  FLAVR from scratch & 0.8216$\pm$0.0306 & 7.03$\pm$3.73 & 0.8083$\pm$0.0421 & 6.96$\pm$2.17 \\
  \textbf{REHRSeg}           & \textbf{0.8306}$\pm$0.0330 & \textbf{5.04}$\pm$2.93 & \textbf{0.8186}$\pm$0.0498 & \textbf{6.57}$\pm$2.38 \\ \bottomrule
  \end{tabular}
  }
  \end{table}

\subsubsection{Comparison of Different Self-SR Methods}
We further assess the impact of replacing our self-SR model with alternative methods for reducing slice thickness. Specifically, we evaluate the B-spline interpolation and SMORE, using their super-resolution  results solely as pseudo-data to assist the segmentation model, as these methods do not provide uncertainty maps or 3D feature maps.
It is evident that B-spline interpolation negatively affects segmentation performance, as the morphology-based interpolation of labels does not accurately reflect the actual boundaries of the B-spline interpolated images, leading to sub-optimal segmentation results.
Although SMORE yields a slight improvement on the DSC scores, they fail to enhance the HD95 metric. In contrast, our method improves both metrics and achieves even better performance when the self-SR model is pretrained on video frame interpolation.

\begin{table}[t]
\centering
\caption{\label{tab:ablation_distillation}
Quantitative results for our ablation study on different knowledge distillation methods.}
\resizebox{\textwidth}{!}{%
\begin{tabular}{lcccc}
\toprule
                        & DSC (LR)      & HD95 (LR) & DSC (HR)      & HD95 (HR) \\ \midrule
MIMIC \citep{MIMIC}             & 0.8202$\pm$0.0319 & 6.13$\pm$3.13 & 0.8085$\pm$0.0485 & 7.03$\pm$2.36 \\
IFVD \citep{IFVD}                 & 0.8269$\pm$0.0270 & 5.90$\pm$2.75 & \textbf{0.8222}$\pm$0.0386 & 6.75$\pm$2.08 \\
Correlation Only             & 0.8321$\pm$0.0278 & 5.98$\pm$3.39 & 0.8048$\pm$0.0533 & 6.75$\pm$2.46 \\ \midrule
REHRSeg ($\lambda$=0.01) & 0.8234$\pm$0.0332 & 7.99$\pm$6.19 & 0.8076$\pm$0.0514 & 6.89$\pm$2.26 \\
REHRSeg ($\lambda$=0.1)  & \textbf{0.8327}$\pm$0.0355 & 5.29$\pm$2.55 & 0.8164$\pm$0.0576 & 6.77$\pm$2.27 \\
\textbf{REHRSeg ($\boldsymbol{\lambda}$=1.0)}  & 0.8306$\pm$0.0330 & \textbf{5.04}$\pm$2.93 & 0.8186$\pm$0.0498 & \textbf{6.57}$\pm$2.38 \\
REHRSeg ($\lambda$=10.0) & 0.8200$\pm$0.0500 & 6.16$\pm$3.19 & 0.8160$\pm$0.0472 & 7.01$\pm$2.35 \\ \bottomrule
\end{tabular}
}
\end{table}

\subsubsection{Comparison of Different Knowledge Distillation Methods}
We also evaluate the effect of incorporating different knowledge distillation methods in REHRSeg.
For comparison, we include the established distillation techniques for dense prediction tasks: MIMIC \citep{MIMIC} with pixel-wise distillation, IFVD with cosine similarity distillation, and REHRSeg using only correlation distillation (\emph{i.e.}, without cosine similarity).
We further evaluate the effect of changing the distillation intensity, which is reflected by the weight of distillation loss during training.
Quantitative results in Table \ref{tab:ablation_distillation} demonstrate that all methods effectively distill some useful knowledge from self-SR model, especially for the HR segmentation compared with the third row in Table \ref{tab:ablation_structure}. 
The cosine similarity and correlation-based distillation methods are generally better than pixel-wise distillation using MSE loss, while the proposed distillation strategy using the combination of these methods can further improve the performance.
We also find that setting the distillation intensity $\lambda=1.0$ yields optimal overall results, with too large deviations from this value resulting in performance degradation.
% This can be achieved by setting the distillation intensity to an appropriate value.

\section{Conclusion and Discussion}
\label{sec:conclusion}
HR MRI segmentation has been extensively studied to provide accurate and detailed delineation of ROI, while encountering real-world application issues with only LR images in hand.
Although existing methods have explored using LR images as inputs for HR segmentation in the inference stage, they still require HR images and their annotations for training, which are both expensive to acquire in clinical scenarios.
In this paper, we propose a novel resource-efficient 3D HR segmentation framework, REHRSeg, that can be trained without the need of HR data.
By further investigating the self-supervised super-resolution framework, we take advance of its capacities in enhancing the segmentation model from three different aspects.
Experimental results for both the synthetic and real-world datasets are promising, as REHRSeg can obtain high-quality HR segmentation, while also achieving superior LR segmentation results in conventional settings.

Despite the inspiring results as mentioned above, there are still potential extensions that can be explored. For example, more advanced uncertainty-guided strategy \citep{uncertainty_guidance_extend} can be used to refine the predicted ROI boundaries, and deeper integration of feature interaction between super-resolution and segmentation \citep{DS2F} may further improve the performance, albeit with an increase in computational cost.
Moreover, the proposed REHRSeg method is anticipated to enhance other segmentation protocols such as few-shot learning and domain adaptation, which can further contribute to the developments of resource-efficient medical image segmentation.

\bibliographystyle{elsarticle-num} 
\bibliography{ref}

\end{document}